\documentclass[twocolumn]{IEEEtran}
\IEEEoverridecommandlockouts
\usepackage{amsmath,amssymb,amsfonts}
\usepackage{algorithmic}
\usepackage{graphicx}
\usepackage{textcomp}
\def\BibTeX{{\rm B\kern-.05em{\sc i\kern-.025em b}\kern-.08em
    T\kern-.1667em\lower.7ex\hbox{E}\kern-.125emX}}

\usepackage{amsthm}
\usepackage[url,USenglish,hyperrefblack,notheorems,IEEEtran]{research17}
\usepackage{lipsum} 
 \usepackage{balance}
\usepackage{booktabs} 
\usepackage{enumitem}
\usepackage{mathtools}
\usepackage[lined,boxed,commentsnumbered,linesnumbered,ruled]{algorithm2e}
\usepackage{tikz} 
\usetikzlibrary{shapes, patterns,decorations.text, decorations.pathreplacing}

\newtheorem{lemma}{\mylemmaname}
\newtheorem{theorem}{\mytheoremname}
\newtheorem{definition}{\mydefinitionname}
\newtheorem{proposition}{\mypropositionname}
\newtheorem{corollary}{\mycorollaryname}
\newtheorem{example}{\myexamplename}

\newtheorem{claim}{\myclaimname}

\usepackage[capitalize]{cleveref}
\crefname{equation}{\unskip}{\unskip}
\crefname{claim}{Claim}{Claims} 
\usepackage{array}
\usepackage{multirow}
\newcolumntype{C}[1]{>{\centering\arraybackslash}p{#1}}
\allowdisplaybreaks

\renewcommand{\vect}[1]{\vectg{#1}} 
\renewcommand{\mat}[1]{\bm{#1}} 
\renewcommand{\vmat}[1]{\bm{#1}} 
\newcommand{\code}[1]{\mathcal{#1}} 
\newcommand{\Nat}[1]{\mathbb{N}_{#1}} 
\newcommand{\collect}[1]{\mathscr{#1}} 
\newcommand*{\Scale}[2][4]{\scalebox{#1}{\ensuremath{#2}}} 
\def\rot#1{\rotatebox{90}{#1}} 
\makeatletter
\renewcommand*\env@matrix[1][*\c@MaxMatrixCols c]{%
  \hskip -\arraycolsep
  \let\@ifnextchar\new@ifnextchar
  \array{#1}}
\makeatother

\renewcommand{\HH}{\mathop{}\!\mathsf{H}}  
\newcommand{\HP}[1]{\HH\left(#1\right)} 
\newcommand{\eHP}[1]{\HH(#1)} 

\newcommand{\BigHP}[1]{\HH\Bigl(#1\Bigr)}
\newcommand{\HPcond}[2]{\HH\left(#1 \kern0.1em\middle|\kern0.1em #2\right)}
\newcommand{\eHPcond}[2]{\HH(#1 \kern0.1em|\kern0.1em #2)} 
\newcommand{\bigHPcond}[2]{\HH\bigl(#1 \kern-0.1em \bigm| \kern-0.1em#2\bigr)}
\newcommand{\BigHPcond}[2]{\HH\Bigl(#1 \kern-0.1em \Bigm| \kern-0.1em#2\Bigr)}
\renewcommand{\II}{\mathop{}\!\mathsf{I}}  
\newcommand{\MI}[2]{\II\left(#1 \kern0.1em{;}\kern0.1em #2\right)} 
\newcommand{\eMI}[2]{\II(#1 \kern0.1em{;}\kern0.1em #2)} 
\newcommand{\bigMI}[2]{\II\bigl(#1 \kern0.1em{;}\kern0.1em #2\bigr)}
\newcommand{\BigMI}[2]{\II\Bigl(#1 \kern0.1em{;}\kern0.1em #2\Bigr)}
\newcommand{\MIcond}[3]{\II\left(#1 \kern0.1em{;}\kern0.1em #2 \kern0.1em\middle|\kern0.1em #3\right)}
\newcommand{\eMIcond}[3]{\II(#1 \kern0.1em{;}\kern0.1em #2 \kern0.1em|\kern0.1em #3)} 
\newcommand{\bigMIcond}[3]{\II\bigl(#1 \kern0.1em{;}\kern0.1em #2 \kern-0.1em \bigm| \kern-0.1em#3\bigr)}
\newcommand{\BigMIcond}[3]{\II\Bigl(#1 \kern0.1em{;}\kern0.1em #2 \kern-0.1em \Bigm| \kern-0.1em#3\Bigr)}
\renewcommand{\r}{\color{black}} 
\renewcommand{\b}{\color{black}} 

\begin{document}

\title{On the Fundamental Limit of Private Information Retrieval for Coded Distributed Storage}

   \author{Hsuan-Yin~Lin,~\IEEEmembership{Member,~IEEE},
  Siddhartha~Kumar,~\IEEEmembership{Student~Member,~IEEE},\\
  Eirik~Rosnes,~\IEEEmembership{Senior~Member,~IEEE}, and
  Alexandre~Graell~i~Amat,~\IEEEmembership{Senior~Member,~IEEE}
  \thanks{This work was partially funded by the Research Council of Norway (grant 240985/F20) and the Swedish Research
    Council (grant \#2016-04253). This paper was presented in part at the IEEE Information Theory Workshop (ITW), Guangzhou, China, November 2018.}%
  \thanks{H.-Y.\ Lin, S.\ Kumar, and E.\ Rosnes are with Simula UiB, N-5020 Bergen, Norway (e-mail: kumarsi@simula.no;
    lin@simula.no; eirikrosnes@simula.no).}  \thanks{A.\ Graell i Amat is with the Department of Electrical Engineering,
    Chalmers University of Technology, SE-41296 Gothenburg, Sweden (e-mail: alexandre.graell@chalmers.se).}  }%


\maketitle

\begin{abstract}
  We consider private information retrieval (PIR) for distributed storage systems (DSSs) with noncolluding nodes where
  data is stored using a non maximum distance separable (MDS) linear code.  It was recently shown that if data is stored
  using a particular class of non-MDS linear codes, the \emph{MDS-PIR capacity}, i.e., the maximum possible PIR rate for
  MDS-coded DSSs, can be achieved.  For this class of codes, we prove that the PIR capacity is indeed equal to the
  MDS-PIR capacity, giving the first family of non-MDS  codes for which the PIR capacity is known.
  For other codes, we provide asymmetric PIR protocols that achieve a strictly larger PIR rate compared to
  existing symmetric PIR protocols.
  %
  %
\end{abstract}

\section{Introduction}

The concept of private information retrieval (PIR) was first introduced by Chor \emph{et al.}
\cite{ChorGoldreichKushilevitzSudan95_1}. A PIR protocol allows a user to privately retrieve an arbitrary data item
stored in multiple servers (referred to as nodes in the sequel) without disclosing any information of the requested item
to the nodes. The efficiency of a PIR protocol is measured in terms of the total communication cost between the user and
the nodes, which is equal to the sum of the upload and download costs. In distributed storage systems (DSSs), data is
encoded by an $[n,k]$ linear code and then stored on $n$ nodes in a distributed manner. Such DSSs are referred to as
coded DSSs \cite{ShahRashmiRamchandran14_1,ChanHoYamamoto15_1}.

One of the primary aims in PIR is the design of efficient PIR protocols from an information-theoretic perspective. Since
the upload cost does not scale with the file size, the download cost dominates the total communication cost
\cite{ChanHoYamamoto15_1,TajeddineElRouayheb16_1}. Thus, the efficiency of a PIR protocol is commonly measured by the
amount of information retrieved per downloaded symbol, referred to as the PIR rate. Recently, Sun and Jafar derived the
maximum achievable PIR rate, the so-called \emph{PIR capacity}, for the case of DSSs with replicated data
\cite{SunJafar17_1, SunJafar18_2}. In the case where the data stored is encoded by an MDS storage code (the so-called
\emph{MDS-coded DSS}) and no nodes collude, a closed-form expression for the PIR capacity, referred to as the
\emph{MDS-PIR capacity}, was derived in \cite{BanawanUlukus18_1}.


In the earlier work \cite{KumarRosnesGraellAmat17_1,KumarLinRosnesGraellAmat17_1sub,LinKumarRosnesGraellAmat18_1}, the
authors focused on the properties of non-MDS storage codes in order to achieve the MDS-PIR capacity. In particular, in
\cite{KumarLinRosnesGraellAmat17_1sub,LinKumarRosnesGraellAmat18_1} it was shown that the MDS-PIR capacity can be
achieved for a special class of non-MDS linear codes, which, with some abuse of language, we refer to as \emph{MDS-PIR
  capacity-achieving} codes (there might exist other codes outside of this class that achieve the MDS-PIR capacity). 
However, it is still unknown whether the
MDS-PIR capacity is the best possible PIR rate that can be achieved for an arbitrarily coded DSS.  In particular, an
expression for the PIR capacity for coded DSSs with arbitrary linear storage codes is still missing.

In this paper, we first prove that the PIR capacity of coded DSSs that use the class of MDS-PIR capacity-achieving codes
introduced in \cite{KumarLinRosnesGraellAmat17_1sub} is equal to the MDS-PIR capacity.  We then address the fundamental
question of what is the maximum achievable PIR rate for an arbitrarily coded DSS. To this purpose, we mainly consider
non-MDS-PIR capacity-achieving codes. Most of the earlier works focus on designing symmetric PIR protocols and it was
shown in \cite{SunJafar17_1,SunJafar17_2,BanawanUlukus18_1} that any PIR scheme can be made symmetric for MDS-coded
DSSs. However, this is in general not the case for non-MDS codes. Specifically, we propose an \emph{asymmetric} PIR
protocol, Protocol~A, that allows asymmetry in the responses from the storage nodes. For non-MDS-PIR capacity-achieving
codes, Protocol~A achieves improved PIR rates compared to the PIR rates of existing symmetric PIR protocols.
Furthermore, we present an asymmetric PIR protocol, named Protocol~B, that applies to non-MDS-PIR capacity-achieving
codes that can be written as a direct sum of MDS-PIR capacity-achieving codes. Finally, we give an example showing that
it is possible to construct an improved (compared to Protocol~A) asymmetric PIR protocol. However, the protocol is
code-dependent and strongly relies on finding \emph{good} punctured MDS-PIR capacity-achieving subcodes of the
non-MDS-PIR capacity-achieving code. 

\section{Preliminaries and System Model}
\label{sec:system-model}

\subsection{Notation and Definitions}

We denote by $\Nat{}$ the set of all positive integers and by $\Nat{a}\eqdef\{1,2,\ldots,a\}$. Vectors are denoted by
lower case bold letters, matrices by upper case bold letters, and sets by calligraphic upper case letters, e.g.,
$\vect{x}$, $\mat{X}$, and $\set{X}$ denote a vector, a matrix, and a set, respectively. In addition, $\comp{\set{X}}$
denotes the complement of a set $\set{X}$ in a universe set. For a given index set $\set{S}$, we also write $X^\set{S}$
and $Y_\set{S}$ to represent $\bigl\{X^{(m)}\colon m\in\set{S}\bigr\}$ and $\bigl\{Y_l\colon l\in\set{S}\bigr\}$,
respectively. The fonts of random and deterministic quantities are not distinguished typographically since it should be
clear from the context.
We denote a submatrix of $\mat{X}$ that is restricted in columns by the set $\set{I}$
by $\mat{X}|_{\set{I}}$. The function $\mathsf{LCM}(n_1,n_2,\ldots,n_a)$ computes the lowest common multiple of $a$
positive integers $n_1,n_2,\ldots,n_a$. The function $\HP{\cdot}$ represents the entropy of its argument and
$\MI{\cdot}{\cdot}$ denotes the mutual information of the first argument with respect to the second
argument.
$\trans{(\cdot)}$ denotes the transpose of its argument. We use the customary code parameters $[n,k]$
to denote a code $\code{C}$ over the finite field $\GF(q)$ of blocklength $n$ and dimension $k$.
A generator matrix of $\code{C}$ is denoted by $\mat{G}^{\code{C}}$, while $\code{C}^{\mat{G}}$ represents the
corresponding code generated by $\mat{G}$. The function $\chi(\vect{x})$ denotes the support of a vector
$\vect{x}$,
while the support of a code $\code{C}$ is defined as the set of coordinates where not all codewords are zero. A
set of coordinates of $\code{C}$, $\set{I}\subseteq\Nat{n}$, of size $k$ is said to be an \emph{information set} if and
only if $\mat{G}^\code{C}|_\set{I}$ is invertible. The $s$-th generalized Hamming weight of an $[n,k]$ code $\code{C}$,
denoted by $d_s^{\code{C}}$, $s\in \Nat{k}$, is defined as the cardinality of the smallest support of an $s$-dimensional
subcode of $\code{C}$.





\subsection{System Model}
\label{sec:system-model}


We consider a DSS that stores $f$ files $\mat{X}^{(1)},\ldots,\mat{X}^{(f)}$, where each file
$\mat{X}^{(m)}=(x_{i,l}^{(m)})$, $m\in\Nat{f}$, can be seen as a random $\beta\times k$ matrix over $\GF(q)$ with
$\beta,k \in\Nat{}$. Assume that each entry $x_{i,l}^{(m)}$ of $\vmat{X}^{(m)}$ is chosen independently and uniformly at
random from $\GF(q)$, $m\in\Nat{f}$. Thus,
\begin{IEEEeqnarray*}{rCl}
  \HP{\vmat{X}^{(m)}}& = &\const{L},\,\forall\,m\in\Nat{f},
  \\
  \HP{\vmat{X}^{(1)},...,\vmat{X}^{(f)}}& = &f\const{L}\quad (\textnormal{in } q\textnormal{-ary units}),
\end{IEEEeqnarray*}
where $\const{L}\eqdef\beta\cdot k$. Each file is encoded using a linear code as follows. Let
$\vect{x}^{(m)}_i=\bigl(x^{(m)}_{i,1},\ldots,x^{(m)}_{i,k}\bigr)$, $i\in\Nat{\beta}$, be a message vector corresponding
to the $i$-th row of $\mat{X}^{(m)}$. Each $\vect{x}^{(m)}_i$ is encoded by an $[n,k]$ code $\code{C}$ over $\GF(q)$
into a length-$n$ codeword $\vect{c}^{(m)}_i=\bigl(c^{(m)}_{i,1},\ldots,c^{(m)}_{i,n}\bigr)$. The $\beta f$ generated
codewords $\vect{c}_i^{(m)}$ are then arranged in the array
$\mat{C}=\trans{\bigl(\trans{(\mat{C}^{(1)})}|\ldots|\trans{(\mat{C}^{(f)})}\bigr)}$ of dimensions $\beta f \times n$,
where $\mat{C}^{(m)}=\trans{\bigl(\trans{(\vect{c}^{(m)}_1)}|\ldots|\trans{(\vect{c}^{(m)}_{\beta})}\bigr)}$. The code
symbols $c_{1,l}^{(m)},\ldots,c_{\beta,l}^{(m)}$, $m\in\Nat{f}$, for all $f$ files are stored on the $l$-th storage
node, $l\in\Nat{n}$.

\subsection{Privacy Model}
\label{sec:privacy}


To retrieve file $\mat{X}^{(m)}$ from the DSS, the user sends a random query $Q_l^{(m)}$ to the $l$-th node for all
$l\in\Nat{n}$. In response to the received query, node $l$ sends the response $A^{(m)}_l$ back to the user. $A^{(m)}_l$
is a deterministic function of $Q_l^{(m)}$ and the code symbols stored in the node.
\begin{definition}
  \label{Def:perfect-PIR}
  Consider a DSS with $n$ noncolluding nodes storing $f$ files. A user who wishes to retrieve the $m$-th file sends the
  queries {\b{$Q^{(m)}_l$}}, $l\in\Nat{n}$, to the storage nodes, which return the responses $A^{(m)}_l$. This scheme
  achieves perfect information-theoretic PIR if and only if
  \begin{IEEEeqnarray}{rCl}
    \IEEEeqnarraymulticol{3}{l}{%
      \text{Privacy:} }\nonumber\\*\quad%
    && \bigMI{m}{Q^{(m)}_l,A^{(m)}_l,\vmat{X}^{(1)},\ldots,\vmat{X}^{(f)}}=0,\,\forall\,l\in\Nat{n},
    \IEEEyesnumber\IEEEyessubnumber\IEEEeqnarraynumspace\label{eq:privacy}
    \\
    \IEEEeqnarraymulticol{3}{l}{%
      \text{Recovery:} }\nonumber\\*\quad%
    && \bigHPcond{\vmat{X}^{(m)}}{A^{(m)}_1,\ldots,A^{(m)}_n,Q^{(m)}_1,\ldots,Q^{(m)}_n}=0.
    \IEEEyessubnumber\label{eq:recovery}
  \end{IEEEeqnarray}
\end{definition}




\subsection{PIR Rate and Capacity}
\begin{definition}
  \label{def:def_PIRrate}
  The PIR rate of a PIR protocol, denoted by $\const{R}$, is the amount of information retrieved per downloaded symbol,
  i.e., $\const{R}\eqdef\frac{\beta k}{\const{D}}$, where $\const{D}$ is the total number of downloaded symbols for the
  retrieval of a single file.
\end{definition}

We will write $\const{R}(\mathcal C)$ to highlight that the PIR rate depends on the underlying storage code
$\code{C}$.
%
It was shown in \cite{BanawanUlukus18_1} that for the noncolluding case and for a given number of
files $f$ stored using an $[n,k]$ MDS code, the MDS-PIR capacity  is
\begin{equation}
  \const{C}^{[n,k]}_f\eqdef\frac{n-k}{n}\inv{\left[1-\Bigl(\frac{k}{n}\Bigr)^f\right]},
  \label{eq:PIRcapacity}  
\end{equation}
where superscript “$[n,k]$” indicates the code parameters of the underlying MDS storage code. When the number of files
$f$ tends to infinity, \eqref{eq:PIRcapacity} reduces to
\[\const{C}^{[n,k]}_\infty \eqdef  \lim_{f\to\infty}  \const{C}^{[n,k]}_f = \frac{n-k}{n},\]  
which we refer to as the asymptotic MDS-PIR capacity. Note that for the case of non-MDS linear codes, the PIR capacity
is unknown.




\subsection{MDS-PIR Capacity-Achieving Codes}
\label{sec:PIRachievable-rates_Symmetric}

In \cite{KumarLinRosnesGraellAmat17_1sub}, two symmetric PIR protocols for coded DSSs, named Protocol~1 and Protocol~2,
were proposed and shown to achieve the MDS-PIR capacity for certain important classes of non-MDS codes. Their PIR rates
depend on the following property of the underlying storage code $\code{C}$.
\begin{definition}
  \label{def:PIRachievable-rate-matrix}
  Let $\code{C}$ be an arbitrary $[n,k]$ code. A $\nu\times n$ binary matrix $\mat{\Lambda}_{\kappa,\nu}(\code{C})$ is
  said to be a \emph{PIR achievable rate matrix} for $\code{C}$ if the following conditions are satisfied.
  \begin{enumerate}
  \item \label{item:1} The Hamming weight of each column of $\mat{\Lambda}_{\kappa,\nu}$ is $\kappa$, and
  \item \label{item:2} for each matrix row $\vect{\lambda}_i$, $i\in\Nat{\nu}$, $\chi(\vect{\lambda}_i)$ always contains
    an information set.
  \end{enumerate}
\end{definition}

The following theorem gives the achievable PIR rate of Protocol~1 from \cite[Thm.~1]{KumarLinRosnesGraellAmat17_1sub}.
\begin{theorem}
  \label{thm:PIRachievable-rates_Symmetric}
  Consider a DSS that uses an $[n,k]$ code $\code{C}$ to store $f$ files. If a PIR achievable rate matrix
  $\mat{\Lambda}_{\kappa,\nu}(\code{C})$ exists, then the PIR rate
  \begin{IEEEeqnarray}{rCl}
    \const{R}_{f,\,\mathsf{S}}(\code{C})& \eqdef &
    \frac{(\nu-\kappa)k}{\kappa n}\inv{\left[1-\Bigl(\frac{\kappa}{\nu}\Bigr)^f\right]}
    \label{eq:finitePIRachievable-rate_symmetric}
  \end{IEEEeqnarray}
  is achievable. 
\end{theorem}

In \eqref{eq:finitePIRachievable-rate_symmetric}, we use subscript $\mathsf S$ to indicate that this PIR rate is
achievable by the symmetric Protocol~1 in \cite{KumarLinRosnesGraellAmat17_1sub}. Define
$\const{R}_{\infty,\,\mathsf{S}}(\code{C})$ as the limit of $\const{R}_{f,\,\mathsf{S}}(\code{C})$ as the number of
files $f$ tends to infinity, i.e.,
$\const{R}_{\infty,\,\mathsf{S}}(\code{C})\eqdef \lim_{f\to\infty} \const{R}_{f,\,\mathsf{S}}(\code{C})
=\frac{(\nu-\kappa)k}{\kappa n}$. The asymptotic PIR rate $\const{R}_{\infty,\,\mathsf{S}}(\code{C})$ is also achieved
by the file-independent Protocol~2 from \cite{KumarLinRosnesGraellAmat17_1sub}.


\begin{corollary}
  \label{cor:MDS-PIRcapacity-achieving-matrix}
  If a PIR achievable rate matrix $\mat{\Lambda}_{\kappa,\nu}(\code{C})$ with $\frac{\kappa}{\nu}=\frac{k}{n}$ exists
  for an $[n,k]$ code $\code{C}$, then the MDS-PIR capacity \eqref{eq:PIRcapacity} is achievable.
\end{corollary}

\begin{definition}
  \label{def:MDS-PIRcapacity-achieving-codes}
  A PIR achievable rate matrix $\mat{\Lambda}_{\kappa,\nu}(\code{C})$ with $\frac{\kappa}{\nu}=\frac{k}{n}$ for an
  $[n,k]$ code $\code{C}$ is called an \emph{MDS-PIR capacity-achieving} matrix, and $\code{C}$ is referred to as an
  \emph{MDS-PIR capacity-achieving} code.
\end{definition}

In the following, we briefly state a main result for Protocol~1 and Protocol~2 from \cite{KumarLinRosnesGraellAmat17_1sub} and compare the required number of
stripes and download cost of these protocols.
\begin{theorem}
  \label{rem:PIRdownload_MDS-PIRcapacity-achieving-codes}
  If an MDS-PIR capacity-achieving matrix exists for an $[n,k]$ code $\code{C}$ with $\frac{\kappa}{\nu}=\frac{k}{n}$,
  then the PIR rates $\const{C}_f^{[n,k]}$ and $\const{C}_\infty^{[n,k]}$ are achievable by Protocol~1 and
  Protocol~2 from \cite{KumarLinRosnesGraellAmat17_1sub}, respectively, using the corresponding required $\beta$ and $\const{D}$. From Definition~\ref{def:def_PIRrate}, we
  have
  \begin{IEEEeqnarray}{c}
    \frac{n\const{D}}{\beta}=
    \begin{cases}
      k\inv{\Bigl(\const{C}_f^{[n,k]}\Bigr)} & \textnormal{ for Protocol~1},
      \\[1mm]
      k\inv{\Bigl(\const{C}_\infty^{[n,k]}\Bigr)} & \textnormal{ for Protocol~2}.
    \end{cases}
    \label{eq:relation_beta-Dcost}
  \end{IEEEeqnarray}
  Furthermore, the smallest number of stripes $\beta$ of Protocol~1 and Protocol~2 is equal to $\nu^f$ and
  $\frac{\mathsf{LCM}(k,n-k)}{k}$, respectively.
\end{theorem}

The following theorem from \cite[Thm.~3]{KumarLinRosnesGraellAmat17_1sub} provides a necessary condition for the
existence of an MDS-PIR capacity-achieving matrix.
\begin{theorem}
  \label{thm:general-d_PIRcapacity-achieving-codes}
  If an MDS-PIR capacity-achieving matrix exists for an $[n,k]$ code $\code{C}$, then
  $d_s^{\code{C}}\geq\frac{n}{k}s$, $\forall\,s\in\Nat{k}$.
\end{theorem}


\section{PIR Capacity for MDS-PIR Capacity-Achieving Codes}
\label{sec:PIRcapacity_MDS-PIRcapacity-achieving-odes}


In this section, we prove that the PIR capacity of MDS-PIR capacity-achieving codes is equal to the MDS-PIR capacity.


\begin{theorem}
  \label{thm:converse_MDS-PIRcapacity-achieving-codes}
  Consider a DSS that uses an $[n,k]$ MDS-PIR capacity-achieving code $\code{C}$ to store $f$ files. Then, the maximum
  achievable PIR rate over all possible PIR protocols, i.e., the PIR capacity, is equal to the MDS-PIR capacity
  $\const{C}^{[n,k]}_{f}$ in \eqref{eq:PIRcapacity}.
\end{theorem}
\begin{IEEEproof}
    See Appendix~\ref{sec:converse-proof_coded-PLC}.
\end{IEEEproof}



\Cref{thm:converse_MDS-PIRcapacity-achieving-codes} provides an expression for the PIR capacity for the family of
MDS-PIR capacity-achieving codes (i.e., \eqref{eq:PIRcapacity}). Moreover, for any finite number of files $f$ and in the
asymptotic case where $f$ tends to infinity, the PIR capacity can be achieved using Protocols~1 and 2 from
\cite{KumarLinRosnesGraellAmat17_1sub}, respectively.

\section{Asymmetry Helps: Improved PIR Protocols}
\label{sec:asymmetric-helps}

In this section, we present three asymmetric PIR protocols for non-MDS-PIR capacity-achieving codes, illustrating that
asymmetry helps to improve the PIR rate. By asymmetry we simply mean that the number of symbols downloaded from the
different nodes is not the same, i.e., for any fixed $m\in\Nat{f}$, the entropies $\eHP{A_l^{(m)}}$, $l\in\Nat{n}$, may
be different. This is in contrast to the case of MDS codes, where any asymmetric protocol can be made symmetric while
preserving its PIR rate \cite{SunJafar17_1,SunJafar17_2,BanawanUlukus18_1}.
We start with a simple motivating example showing that the PIR rate of Protocol~1 from
\cite{KumarLinRosnesGraellAmat17_1sub} can be improved for some underlying storage codes.

\subsection{Protocol~1 From \cite{KumarLinRosnesGraellAmat17_1sub} is Not Optimal in General}
\label{sec:NotOptimal_symmetricPIR}

\begin{example}
  \label{ex:PIRrate_bad-code_n5k3}
  Consider the $[5,3]$ code $\code{C}$ with generator matrix
  \begin{equation*}
    \mat{G}=
    \begin{pmatrix}
      1 & 0 & 0 & 1 & 0
      \\
      0 & 1 & 0 & 1 & 0
      \\
      0 & 0 & 1 & 0 & 1
    \end{pmatrix}.
  \end{equation*}
  The smallest possible value of $\frac{\kappa}{\nu}$ for
  which a PIR achievable rate matrix exists is $\frac{2}{3}$ and a corresponding PIR achievable rate matrix is 
   \begin{equation*} 
    \mat{\Lambda}_{2,3}=
    \begin{pmatrix}
      0 & 1 & 1 & 1 & 1
      \\
      1 & 0 & 0 & 1 & 1
      \\
      1 & 1 & 1 & 0 & 0
    \end{pmatrix}.
  \end{equation*}
  It is easy to verify that $\mat{\Lambda}_{2,3}$ above is a PIR achievable rate matrix for code $\code{C}$. Thus, the
  largest PIR rate for $f=2$ files with Protocol~1 from \cite{KumarLinRosnesGraellAmat17_1sub} is
  $\const{R}_{2,\,\mathsf{S}}=\frac{3^3}{5\cdot 10}=\frac{27}{50}$. In
  Table~\ref{tab:nonMDS-PIRcapacity-achieving-code_n5k3} (taken from \cite[Sec.~IV]{KumarLinRosnesGraellAmat17_1sub}),
  we list the downloaded sums of code symbols when retrieving file $\mat{X}^{(1)}$ and $f=2$ files are stored. In the
  table, for each $m\in\Nat{2}$ and $\beta=\nu^f=3^2$, the interleaved code array $\mat{Y}^{(m)}$ with row vectors
  $\vect{y}^{(m)}_i=\vect{c}^{(m)}_{\pi(i)}$, $i\in\Nat{3^2}$, is generated (according to Protocol~1 from
  \cite{KumarLinRosnesGraellAmat17_1sub}) by a randomly selected permutation function $\pi(\cdot)$.
  \begin{table*}[htbp]
    \centering
    \caption{Protocol~1 with a $[5,3]$ non-MDS-PIR capacity-achieving code for $f=2$}
    \label{tab:nonMDS-PIRcapacity-achieving-code_n5k3}
    \vspace{-7.5mm}
    \begin{IEEEeqnarray*}{rCl}
      \begin{IEEEeqnarraybox}[
        \IEEEeqnarraystrutmode
        \IEEEeqnarraystrutsizeadd{3pt}{1pt}]{v/c/V/c/v/c/v/c/v/c/v/c/v/c/v}
        \IEEEeqnarrayrulerow\\
        & && && \text{Node } 1  && \text{Node } 2 &&
        \text{Node } 3 && \text{Node } 4 &&
        \text{Node } 5 &\\
        \hline\hline
        & && && y^{(1)}_{2({\r 2}-1)+1,1} &&  y^{(1)}_{2({\r 1}-1)+1,2} && y^{(1)}_{2({\r 1}-1)+1,3} &&
        y^{(1)}_{2({\r 1}-1)+1,4} && y^{(1)}_{2({\r 1}-1)+1,5} &
        \\
        & && && y^{(1)}_{2({\r 2}-1)+2,1} &&  y^{(1)}_{2({\r 1}-1)+2,2} && y^{(1)}_{2({\r 1}-1)+2,3} &&
        y^{(1)}_{2({\r 1}-1)+2,4} && y^{(1)}_{2({\r 1}-1)+2,5} &
        \\*\cline{5-15}
        & && \rot{\rlap{\text{round} 1}}&&y^{(2)}_{3\cdot 0+{\r 2},1} &&  y^{(2)}_{3\cdot 0+{\r 1},2}
        && y^{(2)}_{5\cdot 0+{\r 1},3}  &&y^{(2)}_{3\cdot 0+{\r 1},4} && y^{(2)}_{3\cdot 0+{\r 1},5} &
        \\
        & \rot{\rlap{\text{repetition} 1}}&& &&y^{(2)}_{3\cdot 0+{\r 3},1} && y^{(2)}_{3\cdot 0+{\r 3},2}
        && y^{(2)}_{3\cdot 0+{\r 3},3}  &&y^{(2)}_{3\cdot 0+{\r 2},4} && y^{(2)}_{3\cdot 0+{\r 2},5}
        \\*\cline{3-15}
        & && \text{rnd}.~2 &&y^{(1)}_{2\cdot 3+{\r 2},1}+y^{(2)}_{3\cdot 0+{\b 1},1} &&
        y^{(1)}_{2\cdot 3+{\r 1},2}+y^{(2)}_{3\cdot 0+{\b 2},2}
        && y^{(1)}_{2\cdot 3+{\r 1},3}+y^{(2)}_{3\cdot 0+{\b 2},3} &&
        y^{(1)}_{2\cdot 3+{\r 1},4}+y^{(2)}_{3\cdot 0+{\b 3},4}&& y^{(1)}_{2\cdot 3+{\r 1},5}+y^{(2)}_{3\cdot 0+{\b 3},5} &
        \\*\hline\hline
        & && && y^{(1)}_{2({\r 3}-1)+1,1} &&  y^{(1)}_{2({\r 3}-1)+1,2} && y^{(1)}_{2({\r 3}-1)+1,3} &&
        y^{(1)}_{2({\r 2}-1)+1,4} && y^{(1)}_{2({\r 2}-1)+1,5} &
        \\
        & && && y^{(1)}_{2({\r 3}-1)+2,1} &&  y^{(1)}_{2({\r 3}-1)+2,2} && y^{(1)}_{2({\r 3}-1)+2,3} &&
        y^{(1)}_{2({\r 2}-1)+2,4} && y^{(1)}_{2({\r 2}-1)+2,5} &
        \\*\cline{5-15}
        & && \rot{\rlap{\text{round} 1}}&&y^{(2)}_{3\cdot 1+{\r 2},1} &&  y^{(2)}_{3\cdot 1+{\r 1},2}
        && y^{(2)}_{3\cdot 1+{\r 1},3}  &&y^{(2)}_{3\cdot 1+{\r 1},4} && y^{(2)}_{3\cdot 1+{\r 1},5} &
        \\
        & \rot{\rlap{\text{repetition} 2}}&& &&y^{(2)}_{3\cdot 1+{\r 3},1} && y^{(2)}_{3\cdot 1+{\r 3},2}
        && y^{(2)}_{3\cdot 1+{\r 3},3}  &&y^{(2)}_{3\cdot 1+{\r 2},4} && y^{(2)}_{3\cdot 1+{\r 2},5}
        \\*\cline{3-15}
        & && \text{rnd}.~2 &&y^{(1)}_{2\cdot 3+{\r 3},1}+y^{(2)}_{3\cdot 1+{\b 1},1} &&
        y^{(1)}_{2\cdot 3+{\r 3},2}+y^{(2)}_{3\cdot 1+{\b 2},2}
        && y^{(1)}_{2\cdot 3+{\r 3},3}+y^{(2)}_{3\cdot 1+{\b 2},3} &&
        y^{(1)}_{2\cdot 3+{\r 2},4}+y^{(2)}_{3\cdot 1+{\b 3},4}&& y^{(1)}_{2\cdot 3+{\r 2},5}+y^{(2)}_{3\cdot 1+{\b 3},5} &
        \\*\IEEEeqnarrayrulerow
      \end{IEEEeqnarraybox}
    \end{IEEEeqnarray*}
    \vskip -4ex 
  \end{table*}    

  
  Observe that since $\{2,3,4\}\subset\chi(\vect{\lambda_1})=\{2,3,4,5\}$ is an information set of $\code{C}$, the five
  sums of
  \begin{IEEEeqnarray*}{c}
    \Scale[0.95]{\bigl\{y^{(1)}_{2({\r 1}-1)+1,5}, y^{(1)}_{2({\r 1}-1)+2,5}, y^{(2)}_{3\cdot 0+{\r 1},5},
      y^{(1)}_{2\cdot 3+{\r 1},5}+y^{(2)}_{3\cdot 0+{\b 3},5}, y^{(2)}_{3\cdot 1+{\r 1},5}\bigr\}}
  \end{IEEEeqnarray*}
  are not necessarily required to recover $\mat{X}^{(1)}$.
  For privacy concerns, notice that the remaining sums of code symbols from the $5$-th node would be
  \begin{IEEEeqnarray*}{c}
    \Scale[0.95]{\bigl\{y^{(2)}_{3\cdot 0+{\r 2},5}, y^{(1)}_{2({\r 2}-1)+1,5}, y^{(1)}_{2\cdot({\r 2}-1)+2,5},
      y^{(2)}_{3\cdot 1+{\r 2},5}, y^{(1)}_{2\cdot 3+{\r 2},5}+y^{(2)}_{3\cdot 1+{\b 3},5}\bigr\}}.
  \end{IEEEeqnarray*}
 This ensures the privacy condition, since for every combination of files, the user downloads the same number of linear
  sums. This shows that by allowing asymmetry in the responses from the storage nodes, the PIR rate can be improved to
  $\frac{27}{50-5}=\frac{27}{45}=\frac{3}{5}$, which is much closer to the MDS-PIR capacity
  $ \const{C}^{[5,3]}_2=\frac{1}{1+\frac{3}{5}}=\frac{5}{8}$.
\end{example}

Example~\ref{ex:PIRrate_bad-code_n5k3} indicates that for a coded DSS using a non-MDS-PIR capacity-achieving code, there
may exist an asymmetric PIR scheme that improves the PIR rate of the symmetric Protocol~1 from
\cite{KumarLinRosnesGraellAmat17_1sub}.

\subsection{Protocol~A: A General Asymmetric PIR Protocol}
\label{sec:PIRachievable-rates_Asymmetric}

In this subsection, we show that for non-MDS-PIR capacity-achieving codes, by discarding the redundant coordinates that
are not required to form an information set within $\chi(\vect{\lambda}_i)$,
$i\in\Nat{\nu}$, it is always possible to obtain a larger PIR rate compared to that of Protocol~1 from
\cite{KumarLinRosnesGraellAmat17_1sub}.

\begin{theorem}
  \label{thm:PIRachievable-rates_Asymmetric}
  Consider a DSS that uses an $[n,k]$  code $\code{C}$ to store $f$ files. If a PIR achievable rate matrix 
  $\mat{\Lambda}_{\kappa,\nu}(\code{C})$ exists, then the PIR rate
  \begin{IEEEeqnarray}{rCl}
    \const{R}_{f,\,\mathsf{A}}(\code{C})& \eqdef &
    \Bigl(1-\frac{\kappa}{\nu}\Bigr)\inv{\left[1-\Bigl(\frac{\kappa}{\nu}\Bigr)^{f}\right]}
    \label{eq:RfA}
  \end{IEEEeqnarray}
  is achievable. 
\end{theorem}
\begin{IEEEproof}
  See Appendix~\ref{sec:proof_ProtocolA}.
\end{IEEEproof}

We will make use of the following lemma from \cite[Lem.~2]{KumarLinRosnesGraellAmat17_1sub}.
\begin{lemma} 
  \label{lem:PIRrate_upper-bound}
  If a matrix $\mat{\Lambda}_{\nu,\kappa}(\mathcal{C})$ exists for an $[n,k]$ code $\mathcal{C}$, then we have
  \begin{IEEEeqnarray*}{rCl}
    \frac{\kappa}{\nu}\geq \frac{k}{n},
  \end{IEEEeqnarray*}
  where equality holds if $\chi(\vect{\lambda}_i)$, $i\in\Nat{\nu}$, are all information sets.
\end{lemma}

Proposition~1 can be easily verified using Lemma~\ref{lem:PIRrate_upper-bound}.

\begin{proposition}
  Consider a DSS that uses an $[n,k]$ code $\code{C}$ to store $f$ files. Then,
  $\const{R}_{f,\,\mathsf{S}}(\code{C})\leq\const{R}_{f,\,\mathsf{A}}(\code{C})\leq\const{C}^{[n,k]}_f$ with equality
  if and only if $\code{C}$ is an MDS-PIR capacity-achieving code.
\end{proposition}

\begin{IEEEproof}
  The result follows since
  \begin{IEEEeqnarray}{rCl}
    \const{R}_{f,\,\mathsf{S}}(\code{C})
    & = &\frac{(\nu-\kappa)k}{\kappa n}\inv{\left[1-\Bigl(\frac{\kappa}{\nu}\Bigr)^f\right]}
    \nonumber\\
    & \leq &\frac{(\nu-\kappa)k}{\kappa n-(\kappa n-\nu k)}\inv{\left[1-\Bigl(\frac{\kappa}{\nu}\Bigr)^{f}\right]}
    \label{eq:use_Lemma2-1}\\
    & = &\Bigl(1-\frac{\kappa}{\nu}\Bigr)\inv{\left[1-\Bigl(\frac{\kappa}{\nu}\Bigr)^{f}\right]}
    =\const{R}_{f,\,\mathsf{A}}(\code{C})
    \nonumber\\
    & = &\inv{\left[1+\frac{\kappa}{\nu}+\cdots+\Bigl(\frac{\kappa}{\nu}\Bigr)^{f-1}\right]}
    \label{eq:use_Lemma2-2}\\
    & \leq &\inv{\left[1+\frac{k}{n} +\cdots+\Bigl(\frac{k}{n}\Bigr)^{f-1}\right]}=\const{C}^{[n,k]}_f,
    \nonumber
  \end{IEEEeqnarray}
  where both \eqref{eq:use_Lemma2-1} and \eqref{eq:use_Lemma2-2} hold since $\frac{\kappa}{\nu}\geq\frac{k}{n}$.
\end{IEEEproof}

In the following, we refer to the asymmetric PIR protocol that achieves the PIR rate in
Theorem~\ref{thm:PIRachievable-rates_Asymmetric} as Protocol~A (thus the subscript $\mathsf A$ in
$\const{R}_{f,\,\mathsf{A}}(\code{C})$ in \eqref{eq:RfA}). Similar to Theorem~\ref{thm:PIRachievable-rates_Symmetric},
there also exists an asymmetric file-independent PIR protocol that achieves the asymptotic PIR rate
$\const{R}_{\infty,\,\mathsf{A}}(\code{C})\eqdef\lim_{f \to \infty}\const{R}_{f,\,\mathsf{A}}(\code{C}) =
1-\frac{\kappa}{\nu}$ and we simply refer to this protocol as the file-independent Protocol~A.\footnote{As for
  Protocol~1 and Protocol~2 from \cite[Remark~2]{KumarLinRosnesGraellAmat17_1sub}}, $\Lambda_{\kappa,\nu}(\code{C})$ can
be used for both the file-dependent Protocol~A and the file-independent Protocol~A.

\subsection{Protocol~B: An Asymmetric PIR Protocol for a Special Class of Non-MDS-PIR Capacity-Achieving Codes}
\label{sec:PIRachievable-rates_DirectSumMPcodes}

In this subsection, we focus on designing an asymmetric PIR protocol, referred to as Protocol~B, for a special class of
$[n,k]$ non-MDS-PIR capacity-achieving codes, where the code is isometric to a \emph{direct sum} of $P\in\Nat{n}$
MDS-PIR capacity-achieving codes \cite[Ch.~2]{macwilliamssloane77_1}. Without loss of generality, we assume that the
generator matrix $\mat{G}$ of an $[n,k]$ non-MDS-PIR capacity-achieving code $\code{C}$ has the structure
\begin{IEEEeqnarray}{rCl}
  \mat{G}& = &
  \begin{pmatrix}
    \mat{G}_1 &           &       &
    \\
              & \mat{G}_2 &       &
    \\
              &           &\ddots & 
    \\
              &           &       &\mat{G}_P
   \end{pmatrix},
   \label{eq:DirectSumCode-assumptions}
\end{IEEEeqnarray}
where $\mat{G}_p$, of size $k_p\times n_p$, is the generator matrix of a punctured MDS-PIR capacity-achieving
subcode $\code{C}^{\mat{G}_p}$, $p\in\Nat{P}$. 

\begin{theorem}
  \label{thm:PIRachievable-rates_DirectSumMPCodes}
  Consider a DSS that uses an $[n,k]$ non-MDS-PIR capacity-achieving code $\code{C}$ to store $f$ files. If the code
  $\code{C}$ is isometric to a direct sum of $P\in\Nat{n}$ MDS-PIR capacity-achieving codes as in
  \eqref{eq:DirectSumCode-assumptions}, then the PIR rate
  \begin{IEEEeqnarray}{rCl}
    \const{R}_{f,\,\mathsf{B}}(\code{C})& \eqdef & \inv{\left(\sum_{p=1}^P
        \frac{k_p}{k}\inv{\Bigl(\const{C}_{f}^{[n_p,k_p]}\Bigr)}\right)}
    \label{eq:PIRrate_fB}
  \end{IEEEeqnarray}
  is achievable. Moreover, the asymptotic PIR rate
  \begin{IEEEeqnarray}{c}
    \const{R}_{\infty,\,\mathsf{B}}(\code{C})\eqdef\lim_{f\to\infty}\const{R}_{f,\,\mathsf{B}}(\code{C}) =
    \inv{\left(\sum_{p=1}^P\frac{k_p}{k}\inv{\Bigl(\const{C}_{\infty}^{[n_p,k_p]}\Bigr)}\right)}
    \label{eq:PIRrate_B}
  \end{IEEEeqnarray}
  is achievable by a file-independent PIR protocol.
\end{theorem}

\begin{IEEEproof}
  See Appendix~\ref{sec:proof_ProtocolB}.
\end{IEEEproof}

\begin{table*}[!t]
     \caption{Responses by Protocol~C with a $[9,5]$ non-MDS-PIR capacity-achieving code}
    \label{tab:nonMDS-PIRcapacity-achieving-code_n9k5}
    \centering
    \def\Hline{\noalign{\hrule height 2\arrayrulewidth}}
    \vskip -2.0ex 
    \begin{tabular}{@{}lccccccccc@{}}
      \Hline \\ [-2.0ex]
       \text{Subresponses}
        &\text{Node } 1 & \text{Node } 2 & \text{Node } 3 & \text{Node } 4 & \text{Node } 5
        & \text{Node } 6& \text{Node } 7 & \text{Node } 8& \text{Node } 9 \\[0.5ex]
      \hline
      \\ [-2.0ex] \hline  \\ [-2.0ex]
      \text{Subresponse} 1 & $I_1+x^{(m)}_{1,1}$ & $I_2$ & $I_3+x^{(m)}_{1,3}$ & $I_4+x^{(m)}_{1,4}$ & $I_5+x^{(m)}_{1,5}$
        & $I_4+I_5$ & $I_3+I_5$ & $I_3+I_4+I_5$ & $I_1+I_2+I_4+I_5$\\
        \text{Subresponse} 2 & $I_6$ & $I_7+x^{(m)}_{1,2}$ & & $I_9$ & $I_{10}$ & & & & $I_6+I_7+I_9+I_{10}$
    \end{tabular}
    \vskip -2ex 
\end{table*}

We remark that Protocol~B requires $\beta=\mathsf{LCM}(\beta_1,\ldots,\beta_P)$ stripes, where $\beta_p$, $p\in\Nat{P}$,
is the smallest number of stripes of either Protocol~1 or Protocol~2 for a DSS that uses only the punctured MDS-PIR
capacity-achieving subcode $\code{C}^{\mat{G}_p}$ to store $f$ files (see the proof in Appendix~\ref{sec:proof_ProtocolB} and 
Theorem~\ref{rem:PIRdownload_MDS-PIRcapacity-achieving-codes} for the smallest number of stripes $\beta_p$).

Theorem~\ref{thm:PIRachievable-rates_DirectSumMPCodes} can be used to obtain a larger PIR rate for the non-MDS-PIR
capacity-achieving code in Example~\ref{ex:PIRrate_bad-code_n5k3}.

\begin{example} \label{ex:ImprovedPIRrate_bad-code-n5k3}
  Continuing with~\cref{ex:PIRrate_bad-code_n5k3}, by elementary matrix operations, the generator matrix of the
  $[5,3]$ code of \cref{ex:PIRrate_bad-code_n5k3} is equivalent to the generator matrix
  \begin{IEEEeqnarray*}{rCl}
    \begin{pmatrix}
      1 & 0 & 1 & 0 & 0
      \\
      0 & 1 & 1 & 0 & 0
      \\
      0 & 0 & 0 & 1 & 1
    \end{pmatrix}
    & = &
    \begin{pmatrix}
      \mat{G}_1 &
      \\
      & \mat{G}_2
    \end{pmatrix}.
  \end{IEEEeqnarray*}
  It can easily be verified that both $\code{C}^{\mat{G}_1}$ and $\code{C}^{\mat{G}_2}$ are MDS-PIR capacity-achieving
  codes. Hence, from Theorem~\ref{thm:PIRachievable-rates_DirectSumMPCodes}, the asymptotic PIR rate
  \begin{IEEEeqnarray*}{rCl}
    \const{R}_{\infty,\,\mathsf{B}}& = &
    \inv{\left(\frac{2}{3}\frac{1}{1-\frac{2}{3}}+\frac{1}{3}\frac{1}{1-\frac{1}{2}}\right)}=\frac{3}{8}
  \end{IEEEeqnarray*}
  is achievable. $\const{R}_{\infty,\,\mathsf{B}}=\frac{3}{8}$ is strictly larger than both
  $\const{R}_{\infty,\,\mathsf{S}}=\frac{3}{10}$ and $\const{R}_{\infty,\,\mathsf{A}}=\frac{1}{3}$. 

\end{example}

\subsection{Protocol~C: Code-Dependent Asymmetric PIR Protocol}
\label{sec:examples_code-dependent-asymmetricPIRprotocol}

In this subsection, we provide a code-dependent, but file-independent asymmetric PIR protocol for non-MDS-PIR
capacity-achieving codes that cannot be decomposed into a direct sum of MDS-PIR capacity-achieving codes as in
\eqref{eq:DirectSumCode-assumptions}. The protocol is tailor-made for each class of storage codes.  
The main principle of the protocol is to further reduce the number of downloaded symbols by looking at punctured MDS-PIR
capacity-achieving subcodes.
Compared to Protocol~A, which is simpler and allows for a closed-form expression for its PIR rate, Protocol~C 
gives larger PIR rates.

The file-independent Protocol~2 from \cite{KumarLinRosnesGraellAmat17_1sub} utilizes \emph{interference symbols}. An
interference symbol can be defined through a summation as \cite{KumarLinRosnesGraellAmat17_1sub}
\begin{IEEEeqnarray*}{c}
  I_{k(h-1)+h'}\eqdef\sum_{m=1}^f\sum_{j=(m-1)\beta+1}^{m\beta} u_{h,j}x^{(m)}_{j-(m-1)\beta,h'},
\end{IEEEeqnarray*}
where $h,h' \in \Nat{k}$ and the symbols $u_{h,j}$ are chosen independently and uniformly at random from the same field
as the code symbols.

\begin{example}
  \label{ex:ImprovedPIRrate_bad-code_n9k5}
  Consider a $[9,5]$ code $\code{C}$ with generator matrix
  \begin{IEEEeqnarray*}{rCl}
    \mat{G}& = &
    \begin{pmatrix}
      1 & 0 & 0 & 0 & 0 & 0 & 0 & 0 & 1
      \\
      0 & 1 & 0 & 0 & 0 & 0 & 0 & 0 & 1 
      \\
      0 & 0 & 1 & 0 & 0 & 0 & 1 & 1 & 0
      \\
      0 & 0 & 0 & 1 & 0 & 1 & 0 & 1 & 1
      \\
      0 & 0 & 0 & 0 & 1 & 1 & 1 & 1 & 1
    \end{pmatrix}.
  \end{IEEEeqnarray*}
  It has $d_2^{\code{C}}=3<\frac{9}{5}\cdot 2$, thus it is not MDS-PIR capacity-achieving (see
  \cref{thm:general-d_PIRcapacity-achieving-codes}). Note that this code cannot be decomposed into a direct sum of
  MDS-PIR capacity-achieving codes as in \eqref{eq:DirectSumCode-assumptions}. 

  The smallest $\frac{\kappa}{\nu}$ for which a PIR achievable rate matrix exists for
  this code is $\frac{2}{3}$, and a corresponding PIR achievable rate matrix is
  \begin{IEEEeqnarray*}{rCl}
    \mat{\Lambda}_{2,3}=
    \begin{pmatrix}
      0 & 1 & 0 & 0 & 0 & 1 & 1 & 1 & 1
      \\
      1 & 0 & 1 & 1 & 1 & 1 & 1 & 1 & 1
      \\
      1 & 1 & 1 & 1 & 1 & 0 & 0 & 0 & 0
    \end{pmatrix}.
  \end{IEEEeqnarray*}
  The idea of the file-independent Protocol~2 from \cite{KumarLinRosnesGraellAmat17_1sub} is to use the information sets
  $\set{I}_1=\{2,6,7,8,9\}$ and $\set{I}_2=\{1,3,4,5,9\}$ to recover the $\beta k = 1\cdot 5$ requested file symbols
  that are located in $\set{I}_3=\{1,2,3,4,5\}$. Specifically, we use the information set $\set{I}_1$ to reconstruct the
  required code symbols located in $\comp{\chi(\vect{\lambda}_1)}=\{1,3,4,5\}$ and
  $\set{I}_2\subseteq\chi(\vect{\lambda}_2)=\{1,3,4,5,6,7,8,9\}$ to reconstruct the required code symbol located in
  $\comp{\chi(\vect{\lambda}_2)}=\{2\}$. Since the code coordinates $\{1,2,4,5,9\}$ form an $[n',k']=[5,4]$ punctured
  MDS-PIR capacity-achieving subcode $\code{C}^{\mat{G}'}$ with generator matrix
  \begin{IEEEeqnarray*}{rCl}
    \mat{G}'& = &
    \begin{pmatrix}
      1 & 0 & 0 & 0 & 1
      \\
      0 & 1 & 0 & 0 & 1
      \\
      0 & 0 & 1 & 0 & 1
      \\
      0 & 0 & 0 & 1 & 1
    \end{pmatrix},
  \end{IEEEeqnarray*}
  it can be seen that the code coordinates $\{1,4,5,9\}$ are sufficient to correct the erasure located in
  $\comp{\chi(\vect{\lambda}_2)}$. Therefore, compared to Protocol~A, we can further reduce the required number of
  downloaded symbols. The responses from the nodes when retrieving file $\mat{X}^{(m)}$ are listed in
  Table~\ref{tab:nonMDS-PIRcapacity-achieving-code_n9k5}. The PIR rate of Protocol~C is then equal to
  \begin{IEEEeqnarray*}{c}
    \const{R}_{\infty,\,\mathsf{C}}=\frac{1\cdot 5}{n+n'}=\frac{5}{14} < \frac{4}{9}=\const{C}^{[9,5]}_{\infty},
  \end{IEEEeqnarray*}
  which is strictly larger than $\const{R}_{\infty,\,\mathsf{A}}=\frac{1}{3}$. Notice that it can readily
  be seen from Table~\ref{tab:nonMDS-PIRcapacity-achieving-code_n9k5} that the privacy condition in (\ref{eq:privacy}) is
  ensured.

  Finally, we remark that, using the same principle as outlined above, other punctured MDS-PIR capacity-achieving
  subcodes can be used to construct a valid protocol, giving the same PIR rate.  For instance, we could pick the two
  punctured subcodes $\code{C}^{\mat{G}_1}$ and $\code{C}^{\mat{G}_2}$
  with generator matrices
  \begin{IEEEeqnarray*}{rCl}
    \mat{G}_1& = &
    \begin{pmatrix}
      1 & 0 & 0 & 0 & 1 & 1 
      \\
      0 & 1 & 0 & 1 & 0 & 1 
      \\
      0 & 0 & 1 & 1 & 1 & 1 
    \end{pmatrix}\;
    \textnormal{ and }\;
    \mat{G}_2=
    \begin{pmatrix}
      1 & 0 & 1 
      \\
      0 & 1 & 1 
    \end{pmatrix},
  \end{IEEEeqnarray*}
  respectively.
\end{example}

\Cref{ex:ImprovedPIRrate_bad-code_n9k5} above illustrates the main working principle of Protocol~C and how the redundant
set of code coordinates is taken into account. Its general description will be given in a forthcoming extended version.
However, some numerical results are given below, showing that it can attain larger PIR rates than Protocol~A.

\section{Numerical Results}
\label{sec:numerical-results}

In \cref{tab:results}, we compare the PIR rates for different protocols using several binary linear codes. The second
column gives the smallest fraction $\frac{\kappa}{\nu}$ for which a PIR achievable rate matrix exists. In the table,
code $\code{C}_1$ is from Example~\ref{ex:PIRrate_bad-code_n5k3}, code $\code{C}_2$ is from Example~\ref{ex:ImprovedPIRrate_bad-code_n9k5},
$\code{C}_3$ is a $[7,4]$ code with generator matrix $(1,2,4,8,8,14,5)$ (in decimal form, e.g., $\trans{(1,0,1,1)}$ is
represented by $13$) and $d_3^{\code{C}_3}=5<\frac{7}{4}\cdot 3$, and $\code{C}_4$ is an $[11,6]$ code with generator
matrix $(1,2,4,8,16,32,48,40,24,56,55)$ and $d_3^{\code{C}_4}=4<\frac{11}{6}\cdot 3$. Note that $\code{C}_2$,
$\code{C}_3$, and $\code{C}_4$ cannot be decomposed into a direct sum of MDS-PIR capacity-achieving codes as in
\eqref{eq:DirectSumCode-assumptions}. For all presented codes except $\code{C}_3$, Protocol~C achieves strictly larger
PIR rate than Protocol~A, although smaller than the MDS-PIR capacity.

\setlength{\textfloatsep}{18.60004pt plus 2.39996pt minus 4.79993pt}
\begin{table}[!t]
     \caption{PIR rate for different codes and protocols}
    \label{tab:results}
    \centering
    \def\Hline{\noalign{\hrule height 2\arrayrulewidth}}
    \vskip -2.0ex 
    \begin{tabular}{@{}lcccccc@{}}
      \Hline \\ [-2.0ex]
      Code & $\frac{\kappa}{\nu}$ & $\const{R}_{\infty,\,\mathsf{S}}$ & $\const{R}_{\infty,\,\mathsf{A}}$ & $\const{R}_{\infty,\,\mathsf{B}}$
      & $\const{R}_{\infty,\,\mathsf{C}}$ & $\const{C}^{[n.k]}_{\infty}$ \\[0.5ex]
      \hline
      \\ [-2.0ex] \hline  \\ [-2.0ex]
      $\mathcal C_1:[5,3]$  & $2/3$ &$0.3$ & $0.3333$ &  $0.375$ & $0.375$ & $0.4$ \\
      $\mathcal C_2:[9,5]$   & $2/3$ &$0.2778$ & $0.3333$ &  $-$ & $0.3571$ & $0.4444$ \\ 
      $\mathcal C_3:[7,4]$      &   $3/5$                                   &$0.3810$ & $0.4$    &  $-$ & $0.4$    & $0.4286$ \\ 
      $\mathcal C_4:[11,6]$     &    $3/4$                                  &$0.1818$ & $0.25$   & $-$  & $0.2824$ & $0.4545$ \\ 
    \end{tabular}
    \vskip -2ex 
\end{table}

\section{Conclusion}
\label{sec:conclusion}

We proved that the PIR capacity for MDS-PIR capacity-achieving codes is equal to the MDS-PIR
capacity for the case of noncolluding nodes, giving the first family of non-MDS codes for which the PIR capacity
is known. We also showed that allowing asymmetry in the responses from the storage nodes yields larger PIR
rates compared to symmetric protocols in the literature when the storage code is a non-MDS-PIR capacity-achieving
code. We proposed three asymmetric protocols and compared them in terms of PIR rate for different storage codes.


\appendices


\section{Proof of Theorem~\ref{thm:converse_MDS-PIRcapacity-achieving-codes}}
\label{sec:converse-proof_coded-PLC}

Achievability is by Theorem~\ref{thm:PIRachievable-rates_Symmetric} and
Corollary~\ref{cor:MDS-PIRcapacity-achieving-matrix}. Hence, in this appendix, we only provide the converse proof of
Theorem~\ref{thm:converse_MDS-PIRcapacity-achieving-codes}. 
%

Before we proceed with the converse proof, we give some general results that hold for any PIR protocol.

\begin{enumerate}
\item Given a query $Q_l^{(m)}$ sent to the $l$-th node, $m\in\Nat{f}$, the response $A_l^{(m)}$ received by the user is a
  function of $Q_l^{(m)}$ and the $f$ coded chunks (denoted by
  $\vect{c}_l\eqdef\trans{\bigl(c^{(1)}_{1,l},\ldots,c^{(1)}_{\beta,l},c^{(2)}_{1,l},\ldots,c^{(f)}_{\beta,l}\bigr)}$)
  that are stored in the $l$-th node. It follows that
  \begin{IEEEeqnarray}{rCl}
    \bigHPcond{A_l^{(m)}}{Q_l^{(m)},\vmat{X}^{\Nat{f}}}
    & = &\bigHPcond{A_l^{(m)}}{Q_l^{(m)},\vect{c}_l}=0.\IEEEeqnarraynumspace\label{eq:answers}
  \end{IEEEeqnarray}
\item From the condition of privacy, the $l$-th node should not be able to differentiate between the responses $A_l^{(m)}$
  and $A_l^{(m')}$ when the user requests $\vmat{X}^{(m)}$, $m\neq m'$. Hence,
  \begin{IEEEeqnarray}{rCl}
    \bigHPcond{A_l^{(m)}}{\set{Q},\vmat{X}^{(m)}}=\bigHPcond{A_l^{(m')}}{\set{Q},\vmat{X}^{(m)}},
    \label{eq:indistinct-answers}
  \end{IEEEeqnarray}
  where $\set{Q}\eqdef\bigl\{Q^{(m)}_l\colon m\in\Nat{f},\,l\in\Nat{n}\bigr\}$ denotes the set of all possible queries made
  by the user. Although this seems to be intuitively true, a proof of this property is still required and can be found
  in \cite[Lem.~3]{XuZhang17_1sub}.

\item Consider a PIR protocol for a coded DSS that uses an $[n,k]$ code $\code{C}$ to store $f$ files. For any subset of
  files $\set{M}\subseteq\Nat{f}$ and for any information set $\set{I}$ of $\code{C}$, we have
  \begin{IEEEeqnarray}{rCl}
    \bigHPcond{A^{(m)}_\set{I}}{\vmat{X}^{\set{M}},\set{Q}}& = &
    \sum_{l\in\set{I}}\bigHPcond{A^{(m)}_l}{\vmat{X}^{\set{M}},\set{Q}}.
    \label{eq:independent_kAnswers}
  \end{IEEEeqnarray}
  The proof uses the linear independence of the columns of a generator matrix of
  $\code{C}$ corresponding to an information set, and can be seen as a simple extension of \cite[Lem.~2]{BanawanUlukus18_1} or
  \cite[Lem.~4]{XuZhang17_1sub}. 
\end{enumerate}

Next, we state Shearer's Lemma, which represents a very useful entropy method for combinatorial problems.
\begin{lemma}[Shearer's Lemma \cite{Radhakrishnan03_1}]
  \label{lem:Shearer-lemma}
  Let $\collect{S}$ be a collection of subsets of $\Nat{n}$, with each $l\in\Nat{n}$ included in at least $\kappa$ members
  of $\collect{S}$. For random variables $Z_1,\ldots,Z_n$, we have
  \begin{IEEEeqnarray*}{rCl}
    \sum_{\set{S}\in\collect{S}}\eHP{Z_\set{S}}\geq \kappa\eHP{Z_1,\ldots,Z_n}.
  \end{IEEEeqnarray*}
\end{lemma}

Now, we are ready for the converse proof. By {Lemma~\ref{lem:PIRrate_upper-bound}}, since the code
$\code{C}$ is MDS-PIR capacity-achieving, there exist $\nu$ information sets $\set{I}_1,\ldots,\set{I}_\nu$ such that
each coordinate $l\in\Nat{n}$ is included in exactly $\kappa$ members of $\collect{I}=\{\set{I}_1,\ldots,\set{I}_\nu\}$
with $\frac{\kappa}{\nu}=\frac{k}{n}$.

Applying the chain rule of entropy we have
\begin{IEEEeqnarray*}{rCl}
  \bigHPcond{A^{(m)}_{\Nat{n}}}{\vmat{X}^\set{M},\set{Q}}\geq\bigHPcond{A^{(m)}_{\set{I}_i}}{\vmat{X}^\set{M},\set{Q}},
  \quad\forall\,i\in \Nat{\nu}.
\end{IEEEeqnarray*}
  
Let $m\in\set{M}$ and $m'\in\cset{\set{M}}\eqdef\Nat{f}\setminus\set{M}$. Following similar steps as in the proof given in
\cite{BanawanUlukus18_1,XuZhang17_1sub}, we get
\begin{IEEEeqnarray}{rCl}
  \IEEEeqnarraymulticol{3}{l}{%
    \nu\bigHPcond{A^{(m)}_{\Nat{n}}}{\vmat{X}^\set{M},\set{Q}} }\nonumber\\*\quad%
  &\geq &\sum_{i=1}^\nu\bigHPcond{A^{(m)}_{\set{I}_i}}{\vmat{X}^\set{M},\set{Q}}
  \nonumber\\
  & = &\sum_{i=1}^\nu\left(\sum_{l\in\set{I}_i}\bigHPcond{A^{(m)}_l}{\vmat{X}^\set{M},\set{Q}}\right)
  \label{eq:use_kAnswers1}\\[1mm]
  & = &\sum_{i=1}^\nu\left(\sum_{l\in\set{I}_i}\bigHPcond{A^{(m')}_l}{\vmat{X}^\set{M},\set{Q}}\right)
  \label{eq:use_indistinct-answers}\\
  & = &\sum_{i=1}^\nu\bigHPcond{A^{(m')}_{\set{I}_i}}{\vmat{X}^\set{M},\set{Q}}
  \label{eq:use_kAnswers2}\\[1mm]
  & \geq &\kappa\bigHPcond{A^{(m')}_{\Nat{n}}}{\vmat{X}^\set{M},\set{Q}}
  \label{eq:use_Shearer-lemma}\\[1mm]
  & = & \kappa\Bigl[\bigHPcond{A^{(m')}_{\Nat{n}},\vmat{X}^{(m')}}{\vmat{X}^\set{M},\set{Q}}
  \nonumber\\
  && \qquad-\bigHPcond{\vmat{X}^{(m')}}{A^{(m')}_{\Nat{n}},\vmat{X}^\set{M},\set{Q}}\Bigr]
  \nonumber\\
  & = &\kappa\Bigl[\bigHPcond{\vmat{X}^{(m')}}{\vmat{X}^\set{M},\set{Q}}
  \nonumber\\
  && \qquad+\>\bigHPcond{A^{(m')}_{\Nat{n}}}{\vmat{X}^\set{M},\vmat{X}^{(m')},\set{Q}}-0\Bigr]
  \label{eq:use_recovery1}\\
  & = & \kappa\Bigl[\bigHPcond{\vmat{X}^{(m')}}{\vmat{X}^\set{M}}
  \!+\!\bigHPcond{A^{(m')}_{\Nat{n}}}{\vmat{X}^\set{M},\vmat{X}^{(m')},\set{Q}}\Bigr],
  \IEEEeqnarraynumspace\label{eq:use_independence}
\end{IEEEeqnarray}
where \eqref{eq:use_kAnswers1} and \eqref{eq:use_kAnswers2} follow from \eqref{eq:independent_kAnswers},
\eqref{eq:use_indistinct-answers} is because of \eqref{eq:indistinct-answers}, \eqref{eq:use_Shearer-lemma} is due to
Shearer's Lemma, \eqref{eq:use_recovery1} is from the fact that the $m'$-th file $\vmat{X}^{(m')}$ is determined by the
responses $A^{(m')}_{\Nat{n}}$ and the queries $\set{Q}$, and finally, \eqref{eq:use_independence} follows from the
independence between the queries and the files. Therefore, we can conclude that
\begin{IEEEeqnarray}{rCl}
  \IEEEeqnarraymulticol{3}{l}{%
    \bigHPcond{A^{(m)}_{\Nat{n}}}{\vmat{X}^\set{M},\set{Q}}
  }\nonumber\\*%
  & \geq & \frac{\kappa}{\nu}\bigHPcond{\vmat{X}^{(m')}}{\vmat{X}^\set{M}}+\frac{\kappa}{\nu}
  \bigHPcond{A^{(m')}_{\Nat{n}}}{\vmat{X}^\set{M},\vmat{X}^{(m')},\set{Q}}
  \nonumber\\
  & = & \frac{k}{n}
  \bigHPcond{\vmat{X}^{(m')}}{\vmat{X}^\set{M}}
  +\frac{k}{n}\bigHPcond{A^{(m')}_{\Nat{n}}}{\vmat{X}^\set{M},\vmat{X}^{(m')},\set{Q}},
  \IEEEeqnarraynumspace\label{eq:use_MDS-PIRcapaciy-achieving-codes}
\end{IEEEeqnarray}
where we have used Definition~\ref{def:MDS-PIRcapacity-achieving-codes} to obtain
\eqref{eq:use_MDS-PIRcapaciy-achieving-codes}.

Since there are in total $f$ files, we can recursively use \eqref{eq:use_MDS-PIRcapaciy-achieving-codes} $f-1$ times to
obtain
\begin{IEEEeqnarray}{rCl}    
  \IEEEeqnarraymulticol{3}{l}{%
    \bigHPcond{A_{\Nat{n}}^{(1)}}{\vmat{X}^{(1)},\set{Q}} }\nonumber\\*\quad%
  & \geq &\sum_{m=1}^{f-1}\Bigl(\frac{k}{n}\Bigr)^m
  \bigHPcond{\vmat{X}^{(m+1)}}{\vmat{X}^{\Nat{m}}}
  \nonumber\\
  && \qquad+\Bigl(\frac{k}{n}\Bigr)^{f-1}\bigHPcond{A_{\Nat{n}}^{(f)}}{\vmat{X}^{\Nat{f}},\set{Q}}
  \nonumber\\
  & = &\sum_{m=1}^{f-1}\Bigl(\frac{k}{n}\Bigr)^m
  \bigHPcond{\vmat{X}^{(m)}}{\vmat{X}^{\Nat{m-1}}}
  \label{eq:nonegative-entropy}\\
  & = &\sum_{m=1}^{f-1}\Bigl(\frac{k}{n}\Bigr)^m\const{L},\label{eq:use_rank-lemma}
\end{IEEEeqnarray}
where \eqref{eq:nonegative-entropy} follows from \eqref{eq:answers}. 
\eqref{eq:use_rank-lemma} holds since
$\bigHPcond{\vmat{X}^{(m)}}{\vmat{X}^{\Nat{m-1}}}=\HP{\vmat{X}^{(m)}}=\const{L}$.

Now,
\begin{IEEEeqnarray}{rCl}
  \const{L}& = &\HP{\vmat{X}^{(1)}}
  \nonumber\\
  & = &\bigHPcond{\vmat{X}^{(1)}}{\set{Q}}
  -\underbrace{\bigHPcond{\vmat{X}^{(1)}}{A^{(1)}_{\Nat{n}},\set{Q}}}_{=0}
  \label{eq:use_recovery2}\\
  & = &\bigMIcond{\vmat{X}^{(1)}}{A^{(1)}_{\Nat{n}}}{\set{Q}}
  \nonumber\\[1mm]
  & = &\BigHPcond{A^{(1)}_{\Nat{n}}}{\set{Q}}-\BigHPcond{A^{(1)}_{\Nat{n}}}{\vmat{X}^{(1)},\set{Q}}
  \nonumber\\
  & \leq &\BigHPcond{A^{(1)}_{\Nat{n}}}{\set{Q}}-\sum_{m=1}^{f-1}\Bigl(\frac{k}{n}\Bigr)^{m}\const{L},
  \label{eq:use_recursive-step}
\end{IEEEeqnarray}
where \eqref{eq:use_recovery2} follows since any file is independent of the queries $\set{Q}$, and knowing the responses
$A^{(1)}_{\Nat{n}}$ and the queries $\set{Q}$, one can determine $\vmat{X}^{(1)}$. Inequality \eqref{eq:use_recursive-step} holds
because of \eqref{eq:use_rank-lemma}.

Finally, the converse proof is completed by showing that
\begin{IEEEeqnarray}{rCl}
  \const{R}& = &\frac{\const{L}}{\sum_{l=1}^n \BigHP{A^{(1)}_l}}
  \nonumber\\
  & \leq &\frac{\const{L}}{\BigHP{A^{(1)}_{\Nat{n}}}}
  \label{eq:use_chain-rule}\\
  & \leq &\frac{\const{L}}{\BigHPcond{A^{(1)}_{\Nat{n}}}{\set{Q}}}
  \label{eq:use_conditioning-entropy}\\
  & \leq & \frac{1}{1+\sum_{m=1}^{f-1}\bigl(\frac{k}{n}\bigr)^{m}}=\const{C}^{[n,k]}_f,\label{eq:14}
\end{IEEEeqnarray}
where \eqref{eq:use_chain-rule} holds because of the chain rule of entropy, \eqref{eq:use_conditioning-entropy} is due to
the fact that conditioning reduces entropy, and we apply \eqref{eq:use_recursive-step} to obtain \eqref{eq:14}.

\section{Proof of Theorem~\ref{thm:PIRachievable-rates_Asymmetric}}
\label{sec:proof_ProtocolA}
%
%
%
The theorem is proved by showing that some downloaded symbols in Protocol~1 from \cite{KumarLinRosnesGraellAmat17_1sub} are not really necessary both from the recovery and the privacy perspective. The resulting protocol is named Protocol~A, and the proof is based on the fact that for
a PIR achievable rate matrix $\mat{\Lambda}_{\kappa,\nu}(\code{C})$ of a code $\code{C}$, to recover a file of size
$\beta\times k$, exactly  $\nu k$ code coordinates of the $\nu$ information sets
$\{\chi(\vect{\lambda}_i)\}_{i\in\Nat{\nu}}$ are required to be exploited in Protocol~1. In order to illustrate the 
achievability proof, we have to review the steps and proof of Protocol~1 in \cite[Sec.~IV and
App.~B]{KumarLinRosnesGraellAmat17_1sub}, and we refer the reader to \cite{KumarLinRosnesGraellAmat17_1sub} for the details. In particular, Protocol~1 in \cite{KumarLinRosnesGraellAmat17_1sub} is constructed from  two matrices as defined below.
\begin{definition}
  \label{def:PIRinterference-matrices}
  For a given $\nu\times n$ PIR achievable rate matrix $\mat{\Lambda}_{\kappa,\nu}(\code{C})=(\lambda_{u,l})$, we define
  the PIR interference matrices $\mat{A}_{\kappa{\times}n}=(a_{i,l})$ and $\mat{B}_{(\nu-\kappa){\times}n}=(b_{i,l})$
  for the code $\code{C}$ with
  \begin{IEEEeqnarray*}{rCl}
    a_{i,l}& \eqdef &u \text{ if } \lambda_{u,l}=1,\,\forall\,l\in\Nat{n},\,i\in\Nat{\kappa},u\in\Nat{\nu},
    \\
    b_{i,l}& \eqdef &u \text{ if } \lambda_{u,l}=0,\,\forall\,l\in\Nat{n},\,i\in\Nat{\nu-\kappa},u\in\Nat{\nu}.
  \end{IEEEeqnarray*}
\end{definition}

Note that in \cref{def:PIRinterference-matrices}, for each $l\in\Nat{n}$, distinct values of $u\in\Nat{\nu}$ should be
assigned for all $i$. Thus, the assignment is not unique in the sense that the order of the entries of each column of
$\mat{A}$ and $\mat{B}$ can be permuted.
Further, by $\set{S}(a|\mat{A}_{\kappa \times n})$ we denote the set of column coordinates of matrix
$\mat{A}_{\kappa{\times}n}=(a_{i,l})$ in which at least one of its entries is equal to $a$, i.e.,
\begin{IEEEeqnarray*}{rCl}
  \set{S}(a|\mat{A}_{\kappa{\times}n})\eqdef\{l\in\Nat{n}\colon\exists\,a_{i,l}=a,i\in\Nat{\kappa}\}.
\end{IEEEeqnarray*}

Thus, Definition~\ref{def:PIRinterference-matrices} leads to the following claim.
\begin{claim}[{\cite[Claim~1]{KumarLinRosnesGraellAmat17_1sub}}]
  \label{clm:property_PIRinterference-matrices}
  $\set{S}(a|\mat{A}_{\kappa\times n})$ contains an information set of code $\code{C}$,
  $\forall\,a\in\Nat{\nu}$. Moreover, for an arbitrary entry $b_{i,l}$ of $\mat{B}_{(\nu-\kappa)\times n}$,
  $\set{S}(b_{i,l}|\mat{A}_{\kappa\times n})=\set{S}(a|\mat{A}_{\kappa\times n})\subseteq\Nat{n}\setminus\{l\}$ if
  $b_{i,l}=a$.
\end{claim}
From \cref{def:PIRinterference-matrices} we see that there are in total $\kappa n$ entries in $\mat{A}$ and each entry
$a_{i,l}$ is related to a coordinate within $\chi(\vect{\lambda}_i)$, $i\in\Nat{\nu}$, $l\in\Nat{n}$. In Protocol~1 the
user downloads the needed symbols in a total of $\kappa$ repetitions and in the $i$-th repetition, $i\in\Nat{\kappa}$,
the user downloads the required symbols in a total of $f$ rounds. Two types of symbols are downloaded by the user,
\emph{desired symbols}, which are directly related to the requested file (say $\vmat{X}^{(1)}$), and \emph{undesired
  symbols}, which are not related to the requested file, but are exploited to decode the requested file from the desired
symbols.
  
Consider a fixed $i\in\Nat{\kappa}$ and denote by $\const{D}(a_{i,l})$ the total download cost of Protocol~1 resulting
from a particular entry $a_{i,l}$, $l\in\Nat{n}$. First, we focus on the undesired symbols downloaded in
$\mathsf{Step~2}$ of Protocol~1. In each repetition the user downloads
\begin{IEEEeqnarray*}{rCl}
  \IEEEeqnarraymulticol{3}{l}{%
    \kappa\frac{\binom{f-1}{\ell}\bigl[\const{U}(\ell)-1-\const{U}(\ell-1)+1\bigr]}{\kappa}}\nonumber\\*\quad%
  & = &\binom{f-1}{\ell}\kappa^{f-(\ell+1)}(\nu-\kappa)^{\ell-1}
\end{IEEEeqnarray*}
undesired symbols resulting from a particular $a_{i,l}$ in the $\ell$-th round, $\ell\in\Nat{f-1}$, where
$\const{U}(\ell)\eqdef\sum_{h=1}^\ell\kappa^{f-(h+1)}(\nu-\kappa)^{h-1}$. Hence, for the undesired symbols associated
with $a_{i,l}$, in total
\begin{IEEEeqnarray}{rCl}
  \IEEEeqnarraymulticol{3}{l}{%
    \kappa\binom{f-1}{\ell}\kappa^{f-(\ell+1)}(\nu-\kappa)^{\ell-1}
  }\nonumber\\*\quad%
  & = &\binom{f-1}{\ell}\kappa^{f-\ell}(\nu-\kappa)^{\ell-1}
  \label{eq:undesired-symbols_a}
\end{IEEEeqnarray}
symbols are downloaded in every $\ell$-th round of all $\kappa$ repetitions.

Secondly, for a particular entry $a_{i,l}$ in the $i$-th repetition, the user downloads $\kappa^{f-1}$ desired
symbols from the $l$-th node in round $\ell=1$, and
\begin{IEEEeqnarray}{c}
  \const{W}(\ell)-1-\const{W}(\ell-1)+1 =\binom{f-1}{\ell}\kappa^{f-(\ell+1)}(\nu-\kappa)^{\ell}
  \IEEEeqnarraynumspace\label{eq:desired-symbols_a-rndl}
\end{IEEEeqnarray}
extra desired symbols in the $(\ell+1)$-th round, $\ell\in\Nat{f-1}$, where $\const{W}(\ell)$ is defined as
\begin{IEEEeqnarray*}{rCl}
  \const{W}(\ell)& \eqdef &\kappa^{f-1}+\sum_{h=1}^\ell\binom{f-1}{h}\kappa^{f-(h+1)}(\nu-\kappa)^h.
\end{IEEEeqnarray*}

In summary, using \eqref{eq:undesired-symbols_a} and \eqref{eq:desired-symbols_a-rndl}, the download cost associated to entry $a_{i,l}$  is obtained as
\begin{IEEEeqnarray*}{rCl}
  \const{D}(a_{i,l})& = &\sum_{\ell=1}^{f-1}\binom{f-1}{\ell}
  \kappa^{f-\ell}(\nu-\kappa)^{\ell-1}\nonumber\\
  && \>+\sum_{\ell=0}^{f-1}\binom{f-1}{\ell}\kappa^{f-(\ell+1)}(\nu-\kappa)^{\ell}
  \\
  & = &\frac{\nu^f-\kappa^f}{\nu-\kappa}.
\end{IEEEeqnarray*}

In the part of $\mathsf{Step~2}$ of Protocol~1 that exploits side information, we only require $\nu$ information
sets induced by the matrix $\mat{A}$ to reconstruct code symbols induced by $\mat{B}$. Moreover, from
\cite[App.~B]{KumarLinRosnesGraellAmat17_1sub}, after $\mathsf{Step~2}$ of Protocol~1, $\beta=\nu^f$ rows of code
symbols of length $n$ have been downloaded, and again the information sets induced by the matrix $\mat{A}$ are enough to
recover all length-$k$ stripes of the requested file. In other words, $\kappa n-\nu k$ entries of $\mat{A}$ are
redundant for the reconstruction of all $\beta=\nu^f$ stripes of the requested file. Thus, the improved PIR
rate becomes
\begin{IEEEeqnarray*}{rCl}
  \frac{\beta k}{\const{D}}& = &\frac{\nu^f k}{\textnormal{download cost of Protocol~1}
    -(\kappa n-\nu k)\const{D}(a_{i,l})}
  \\
  & = &\frac{\nu^f k}{\frac{\kappa n}{\nu-\kappa}\Bigl[\nu^f-\kappa^f\Bigr]-\frac{\kappa n-\nu
      k}{\nu-\kappa}\Bigl[\nu^f-\kappa^f\Bigr]}
  \\
  & = &\frac{\nu^f k}{\frac{\nu k}{\nu-\kappa}\Bigl[\nu^f-\kappa^f\Bigr]}
  =\Bigl(1-\frac{\kappa}{\nu}\Bigr)\inv{\left[1-\Bigl(\frac{\kappa}{\nu}\Bigr)^{f}\right]}.
\end{IEEEeqnarray*}

Finally, we would like to emphasize that by removing the redundant downloaded sums of code symbols in Protocol~1, it can
be shown that within each storage node in each round $\ell\in\Nat{f}$ of all repetitions, file symmetry still
remains. This follows from a similar argumentation as in the privacy part of the proof of Protocol~1 in
\cite[App.~B]{KumarLinRosnesGraellAmat17_1sub}. In the following, we briefly explain that in each round $\ell\in\Nat{f}$
of all repetitions, for each particular entry $a_{i,l}$ and for every combination of files $\set{M}\subseteq\Nat{f}$
with $\card{\set{M}}=\ell$, the user requests the same number of every possible combination of files in
$\const{D}(a_{i,l})$. 
\begin{itemize}
\item In the first round ($\ell=1$) of all $\kappa$ repetitions, it follows from
\eqref{eq:undesired-symbols_a} that, for each $m'\in\Nat{2:f}$, the number of downloaded undesired symbols resulting from a
particular entry $a_{i,l}$ is $\kappa^{f-1}$, the same as the number of downloaded desired symbols resulting from $a_{i,l}$.
\item In the $(\ell+1)$-th round of all $\kappa$ repetitions,
$\ell\in\Nat{f-2}$, arbitrarily choose a combination of files $\set{M}\subseteq\Nat{2:f}$, where
$\card{\set{M}}=\ell$. For a particular entry $a_{i,l}$, it follows from \eqref{eq:desired-symbols_a-rndl} that the
total number of downloaded desired symbols for files pertaining to $\{1\}\cup\set{M}$ is equal to
$\kappa^{f-(\ell+1)}(\nu-\kappa)^{\ell}$. On the other hand, for the undesired symbols resulting from a particular $a_{i,l}$,
it follows from \eqref{eq:undesired-symbols_a} that in the $(\ell+1)\text{-th}$ round the user downloads
$\kappa^{f-(\ell+1)}(\nu-\kappa)^{\ell}$ linear sums for a combination of files $\set{M}\subseteq\Nat{2:f}$,
$\card{\set{M}}=\ell+1$. Thus, in rounds $\Nat{f-1} \setminus \{1\}$, an equal number of linear sums for all combinations of files
$\set{M}\subseteq\Nat{f}$ are downloaded.
\item In the $f$-th round, only desired symbols are downloaded. Since each desired
symbol is a linear combination of code symbols from all $f$ files, an equal number of linear sums is again downloaded
from each file.
\end{itemize}
In summary, in response to each particular $a_{i,l},$ the user downloads the same number of linear sums for every possible combination of files. As illustrated above, this is inherent from Protocol~1, and hence the privacy condition of \eqref{eq:privacy} is still satisfied.

\section{Proof of Theorem~\ref{thm:PIRachievable-rates_DirectSumMPCodes}}
\label{sec:proof_ProtocolB}

The result 
follows  by treating Protocol~1 and Protocol~2 from \cite{KumarLinRosnesGraellAmat17_1sub} as  subprotocols for
each punctured MDS-PIR capacity-achieving subcode $\code{C}^{\mat{G}_p}$, $p\in\Nat{P}$. 
If Protocol~1 is used as a subprotocol, then we obtain the file-dependent Protocol~B and the PIR rate in \eqref{eq:PIRrate_fB}, while if Protocol~2 is used as a subprotocol, then we obtain the file-independent Protocol~B and the PIR rate in \eqref{eq:PIRrate_B}.

For the asymmetric Protocol~B, we require $\beta=\mathsf{LCM}(\beta_1,\ldots,\beta_P)$ stripes, where $\beta_p$,
$p\in\Nat{P}$, is the smallest number of stripes of either Protocol~1 or Protocol~2 for a DSS that uses only the 
punctured MDS-PIR capacity-achieving subcode $\code{C}^{\mat{G}_p}$ to store $f$ files (see Theorem~\ref{rem:PIRdownload_MDS-PIRcapacity-achieving-codes}). Note that for Protocol~1 the \emph{index
  preparation}\footnote{This terminology was introduced in $\mathsf{Step~1}$ of Protocol~1 from
  \cite{KumarLinRosnesGraellAmat17_1sub}, i.e., the indices of the rows for each file are interleaved randomly and
  independently of each other.} should be made for all $\beta$ stripes. Since $\sum_{p=1}^P k_p=k$ and
$\sum_{p=1}^P n_p=n$, to privately retrieve the entire requested file consisting of $k$ symbols in each stripe, we have to privately recover all $P$ substripes of all $\beta$ stripes, where the $p$-th substripe is of length 
$k_p$, by processing the subprotocol (either Protocol~1 or Protocol~2) for every punctured subcode $\code{C}^{\mat{G}_p}$. In particular,
for each punctured subcode $\code{C}^{\mat{G}_p}$ we repeat the subprotocol $\beta/\beta_p$ times to recover all the 
length-$k_p$ requested substripes. This can be done since both Protocol~1 and Protocol~2 recover $\beta_p$ stripes of length $k_p$,
while repeating it $\beta/\beta_p$ times enables the recovery of $\beta$ length-$k_p$ substripes. Note that privacy is
ensured since the storage nodes of each punctured  subcode are disjoint and within the nodes associated with each punctured subcode
$\code{C}^{\mat{G}_p}$ the subprotocol (Protocol~1 or Protocol~2) yields privacy against each server \cite{KumarLinRosnesGraellAmat17_1sub}.

Denote by $\const{D}_p$ the total download cost for each node for the punctured subcode $\code{C}^{\mat{G}_p}$ using the
subprotocol, $p\in\Nat{P}$. The PIR rates of the file-dependent and file-independent Protocol~B are
given by
\begin{IEEEeqnarray}{rCl}
  \frac{\beta k}{\const{D}}& = &\frac{\beta k}{\sum\limits_{p=1}^P\frac{\beta}{\beta_p}n_p \const{D}_p}
  \label{eq:repeating-subprotocols}
  \\
  & = &\inv{\left(\sum\limits_{p=1}^P\frac{1}{k}\frac{n_p\const{D}_p}{\beta_p}\right)}
  \nonumber\\[1mm]
  & = &
  \begin{cases}
    \inv{\left(\sum\limits_{p=1}^P\frac{1}{k}k_p\inv{\Bigl(\const{C}_f^{[n_p,k_p]}\Bigr)}\right)}
    & \\[2mm]
    &\hspace{-3.75cm}\textnormal{ if Protocol~1 is used as subprotocol,}
    \\[3mm]
    \inv{\left(\sum\limits_{p=1}^P\frac{1}{k}k_p\inv{\Bigl(\const{C}_\infty^{[n_p,k_p]}\Bigr)}\right)}
    & \\[2mm]
    &\hspace{-3.75cm}\textnormal{ if Protocol~2 is used as subprotocol},
  \end{cases}
  \IEEEeqnarraynumspace\label{eq:use_beta-Dcost-relation}
\end{IEEEeqnarray}
where \eqref{eq:repeating-subprotocols} holds since within each punctured subcode, the subprotocol is required to be repeated 
$\frac{\beta}{\beta_p}$ times and \eqref{eq:use_beta-Dcost-relation} follows from \eqref{eq:relation_beta-Dcost}.

\IEEEtriggeratref{2}


\bibliographystyle{IEEEtran}
\bibliography{./defshort1,./biblio1}

\end{document}